# The Integration of YBCO Coated Conductors into Magnets and Rotating Machinery

G. A. Levin and P. N. Barnes

Propulsion Directorate, Air Force Research Laboratory, 1950 Fifth St. Bldg. 450, Wright-Patterson Air Force Base OH 45433

## ABSTRACT

The implementation of the $2^{nd}$ generation high-$T_c$ superconductors in power applications, such as electrical transformers, motors and generators requires superconducting wires that are superior to copper Litz wires at cryogenic temperatures in terms of losses in time-varying magnetic field, as well as in engineering current density. Another problem is to find a way to make practical coils and armatures out of flat tape-like conductors with low bending strain tolerance. We discuss several novel approaches to the construction of coils and armatures based specifically on the properties of coated conductors manufactured today.

## INTRODUCTION

Two major shortcomings of YBCO coated conductors need to be overcome in order to facilitate their implementation in ac applications, such as transformers and armature winding of motors and generators [1]. One issue associated with the coated conductors, which are manufactured in the form of thin and relatively wide tapes, is the high hysteresis loss in time-varying magnetic field. The other is their attendant mechanical properties that are very different from the traditional material such as copper Litz wire. Bending strain limitations restrict the types of windings configurations that are possible compared to copper.

The main route to hysteresis (and overall) loss reduction that has been explored in recent years is the replacement of the uniform wide YBCO film with a set of parallel narrow filaments (stripes)[2-8]. The first experiments have demonstrated that the hysteresis loss in experimental multifilamentary samples can be reduced by at least an order of magnitude. However, the coupling loss in the multifilamentary coated conductors can become comparable to the hysteresis loss at a sweep rate B$f$ of a few Tesla/s when the twist pitch is equal to 20 cm (B is the amplitude of the magnetic field and $f$ is the frequency)[6]. Therefore, more work is needed in order to achieve a substantial – one or two orders of magnitude – reduction of total losses (hysteresis and coupling) at the operating sweep rate of at least 10 T/s.

The second shortcoming of coated conductors is their low tolerance to bending and twisting strain. This requires an almost complete reexamination of the winding technique. The problem of ac losses and the mechanical properties of the conductor becomes intertwined because the twisting of the multifilamentary conductor is necessary in order to limit coupling losses. Here we present several novel approaches to making a twisted conductor and wiring coils and armatures with the 2$^{nd}$ generation YBCO coated conductors.

## DOUBLE PANCAKE COILS.

The double pancake coils is a preferred form of making the field windings in rotating machinery because both ends of the coil are on the outside as opposed to a simple pancake coil [9]. There are indications, however, that the sideways bending of the innermost turn degrades the current-carrying capacity of the 1$^{st}$ generation wires (Bi-2223)[10]. Making a conventional double pancake coil out of the 2$^{nd}$ generation wire will prove to be even more problematic because of much higher rigidity of the flat metal tape with respect to the sideways bending.

Figure 1 shows a way to overcome the problem of lateral deformation in flat wide tapes. In Fig. 1(a) a long conductor is cut into two branches with the remaining (uncut) part of the tape allowing the current to flow between the branches (as shown by the arrow). If W and L are respectively the width and length of the initial conductor, the resultant conductor will have approximately a width of W/2 and will be twice as long as the initial piece. The uncut area need only be $W \times W / 2$ to maintain the same critical current as that in both branches.

In Fig. 1(b) a model of a coil former (or mandrel) is shown. In this case, it has a radial slot and two quarter circle (of radius **r**) slots leading to the outer rim of radius **R**. The "lower" part of the conductor shown in Fig. 1(a) is inserted into the slots and each branch is wound in the opposite directions. In such a coil the conductor experiences only the bending strain determined by the radius of curvature **R**, but no lateral strain. A small section of the conductor inside the quarter circle slots (total length π**r**) is subjected to the largest strain as determined by the radius of the slots.

## AC LOSSES AND TWIST PITCH.

In the multifilament coated conductor the total magnetization loss is the sum of losses in the superconducting layer $Q_s$ and in the normal metal $Q_n$ (predominantly the coupling loss). In the limit of full field penetration it is given by[3,6]

$$Q = Q_s + Q_n \approx I_c W_n Bf + k \frac{(BfL)^2}{\rho} d_n W \tag{1}$$

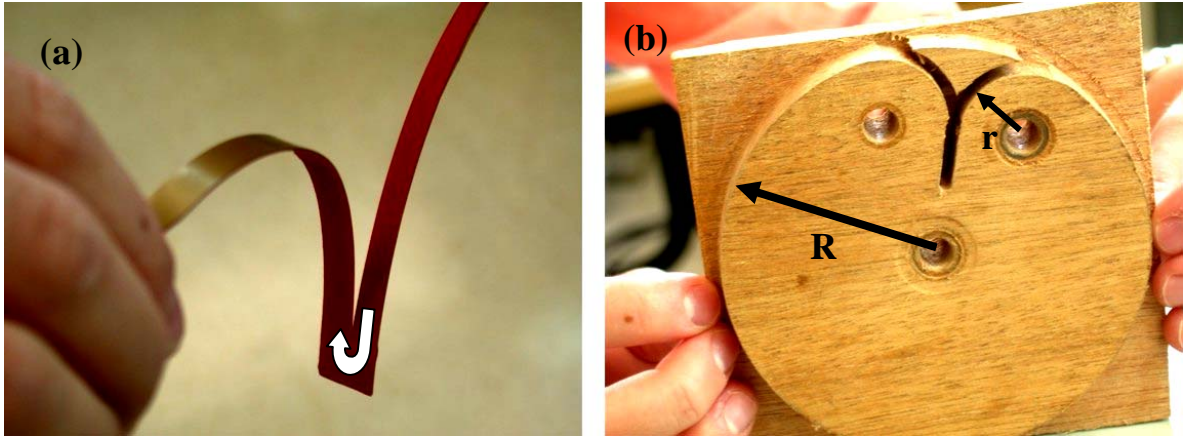

**FIGURE 1**. (a) A model of a tape-like conductor cut along the centerline, but not completely. The remaining part of superconducting layer should allow the supercurrent to flow between the branches. The two branches are bent in the opposite directions. (b) A model of a coil former. The part of the conductor shown in Fig. 1(a) will be inserted into the slots and the two branches wound in the opposite directions.

Here $I_c$ is the critical current, $W_n$ is the width of an individual stripe, $L$ is half of the twist pitch, $\rho$ is the effective resistivity of the substrate, $d_n$ is the thickness of the metal substrate, and $W$ is the width of the conductor. The numerical coefficient $k$ may depend on the type of the twist. The reduction of the coupling loss can be achieved by increasing the effective resistivity and by twisting the conductor. Here we will concentrate on the latter part of this two-prong effort.

Usually in the literature[1,2,11] one can find a description of the axial twist shown in Fig. 2(a). In this case the tape is twisted about its long axis. In Fig. 2(b) we present another option – "bending twist". It is obtained the same way as the conductor shown in Fig. 1(a). In the applied magnetic field the conductors shown in Figs. 2(a) and (b) will expose both of their sides (indicated by darker and brighter tone) thus limiting the amount of magnetic flux passing through the conductor. Each type of twist has its advantages and disadvantages. There are situations where one may be more suitable than the other. In certain situations both types of twist have to be employed in order to achieve the maximum benefit.

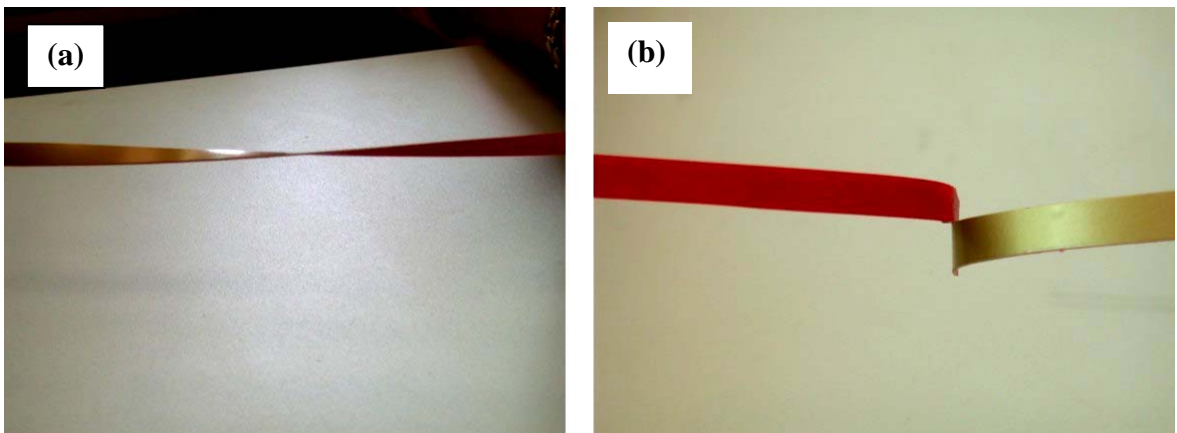

**FIGURE 2**. (a) A conventional axial twist of a tape-like conductor; one side of the tape is darker than the other. (b) bend-twisted tape, view from the top.

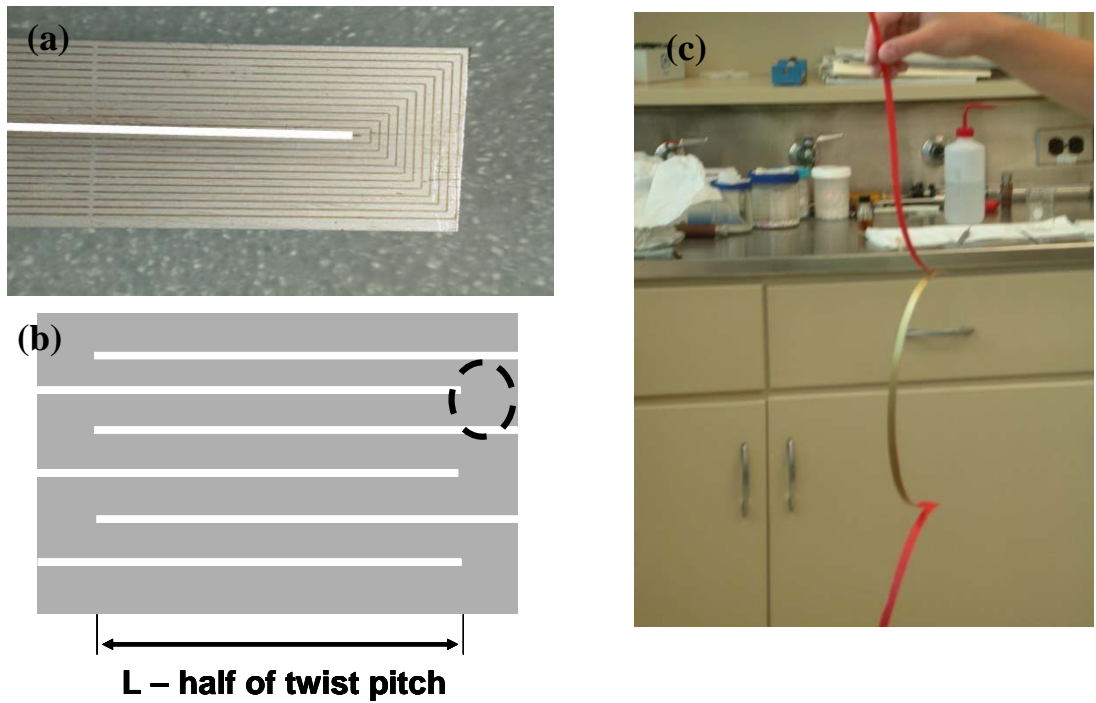

**FIGURE 3**. (a) 12 mm wide coated conductor striated by laser ablation. The stripes have the race-track shape allowing current to flow between the two branches. Subsequently, the conductor is cut along the wide white band in the center. (b) Sketch of a sheet of coated conductor cut in the form of a meander. Unfolded, such a meander becomes bend-twisted conductor with the twist pitch length indicated. (c) A model of unfolded meander conductor.

To reduce the ac losses the conductor has to be made multifilament (striated) and twisted. A type of striation that can be used in conjunction with the bending twist is shown in Fig. 3(a). Here a 12 mm wide conductor was divided into 0.5 mm wide race-track stripes by laser ablation. Then it is cut along the centerline as shown and the two branches bend in the opposite directions forming a twist shown in Fig. 2(b).

**LAYERED COIL CONSRUCTION.**

A conductor that can be bent-twisted may be prepared from a sheet of coated conductor by cutting it into a meander of a desired width as shown in Fig. 3(b). The conductor has to be striated, so that the encircled area in Fig. 3(b) looks like that in Fig. 3(a). A model of unfolded meander conductor is shown in Fig. 3(c).

Since the meander conductor consists of sections offset from each other by the width of the tape, Fig. 4(a), it lends itself naturally to making a solenoidal or layered coil, instead of pancake coil. Figure 4(b) shows a sketch of the cross-section of a coil former (mandrel) for such a coil. The purpose of the radial slot is the same as in the coil former shown in Fig. 1(b). Figure 4(c) shows a model of a layered coil with bent-twisted conductor. In this type of coil, the length of one section of the meander conductor must be equal to the circumference of the mandrel. Thus, each section forms one turn of the coil and each turn is transposed (twisted) by 180° with respect to its neighbors. The result is a coil in which the conductor is twisted with half of the twist pitch equal to the circumference of the coil.

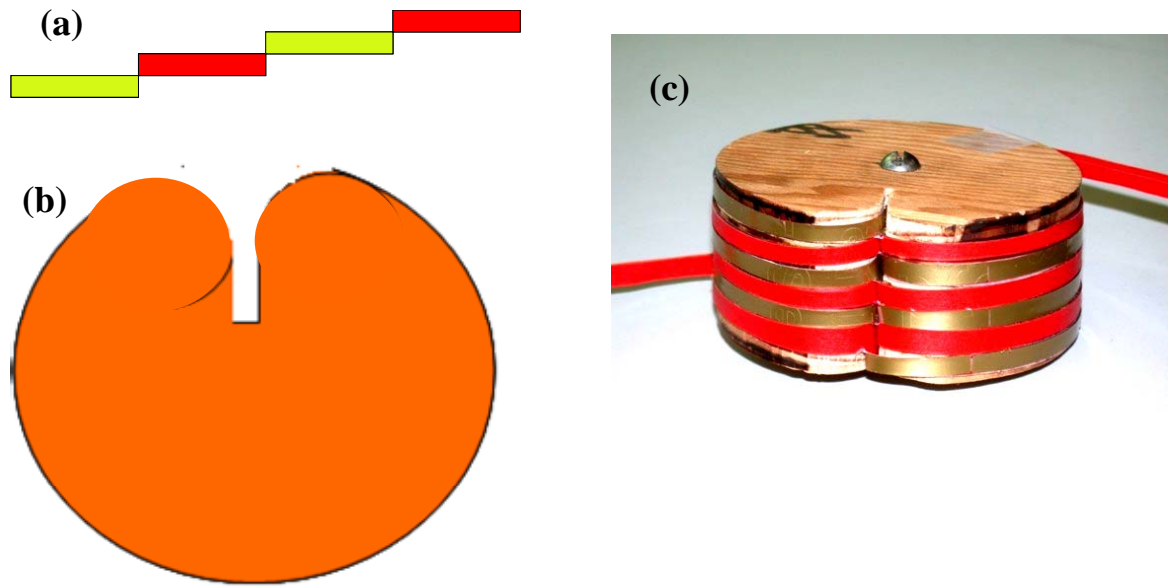

**FIGURE 4.** (a) A sketch of unfolded meander conductor. The darker and brighter sections correspond to the "front" and "backside" of the tape. (b) A sketch of the cross-section of the coil former. (c) A model of helical (layered) coil. Notice that each turn is transposed (twisted) by 180° with respect to its neighbors. The darker and brighter tone corresponds to the opposite sides of the tape.

In order to increase the current carrying capacity of the conductor, several sheets like the one shown in Fig. 3(b) can be stacked on top of each other and then unfolded as shown in Fig.3(c), thus making it a multi-ply twisted conductor. It is also possible to make a coil similar to that in Fig.4(c) with several layers. However, it is desirable to limit the number of layers and it is preferable to use coils like that in Fig. 4(c) as high current, low voltage coils such as the secondary coil in step-down transformers.

## CONSRUCTION OF STATOR WINDINGS.

There are several types of armature windings for motors and generators[12,13]. The shapes and the methods of winding of traditional armatures are determined in large measure by the mechanical properties of copper conductors. Substantial revision of these approaches is required if we want to make the stator windings out of tape-like conductors. Here we will consider as examples the Cramme ring armature and diamond-shaped coils.

### Gramme Ring Armature

This type of wiring is less efficient than the diamond-shaped armature, but it has been suggested recently as the type suitable for all-cryogenic aircraft generators[11]. The Gramme ring wiring is somewhat similar to a helical coil. Using meander conductor like the one shown in Figs. 3(b,c), a Gramme ring coil can be wound as shown in Figs. 5(a,b). The active length of the conductor can be placed in the slots edgewise, which may be beneficial for loss reduction because such orientation minimizes the component of the magnetic field perpendicular to the wide face of the tape. The return path is shown with the wide face normal to the magnetic field of the rotor shown by the arrow. If necessary, the return part can also be placed in the slots edgewise. The coil shown in Fig. 5 incorporates both types of twist – axial and bending. An essential advantage of the bending twist is that the direction of the current can be changed by almost $180^\circ$ over very small distance, comparable to the width of the conductor.

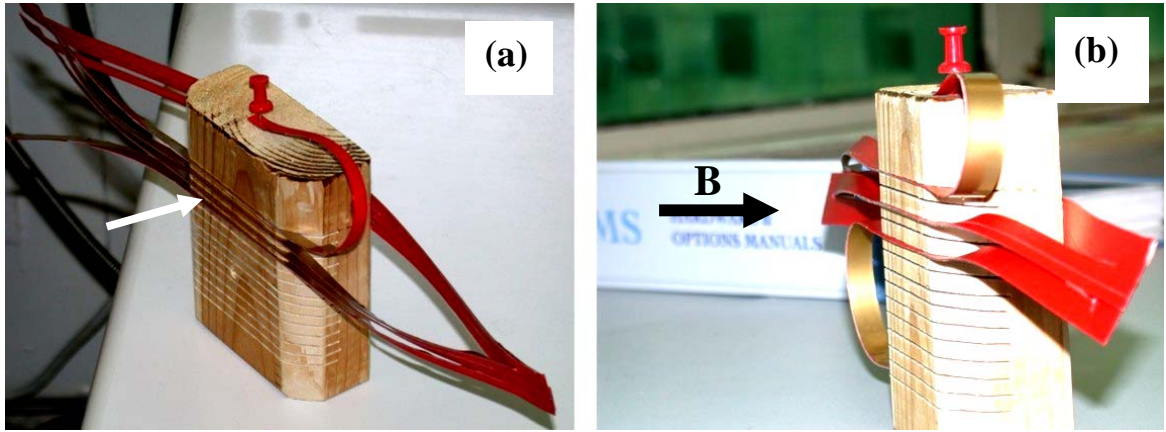

**FIGURE 5.** (a) An overall view of a model Gramme ring-type wiring. The white arrow indicates the active length. (b) A close-up view of the same coil. The active length of the conductor is placed edgewise into the slots as to minimize the component of the magnetic field perpendicular to the wide face of the tape. The direction of the magnetic field at its maximum is indicated by the black arrow. Notice that the tape is axially twisted by approximately $45^\circ$.

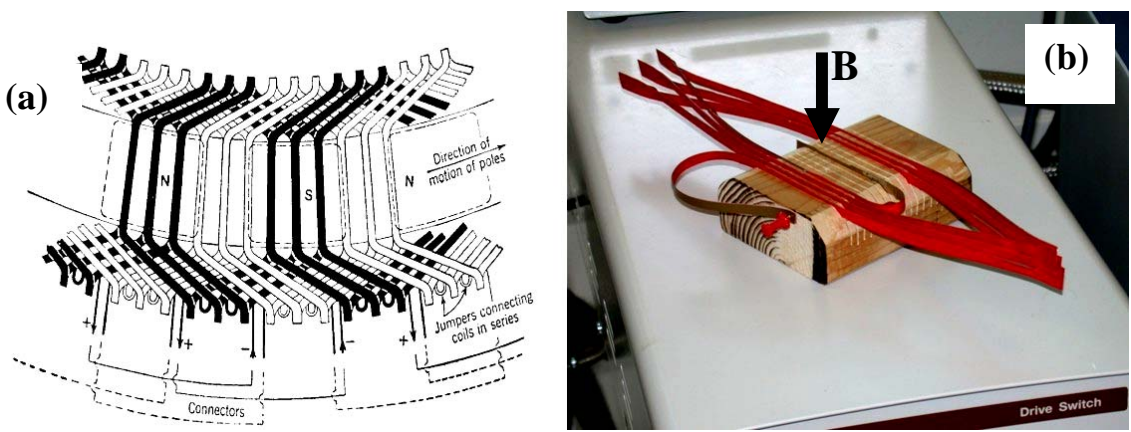

**FIGURE 6.** (a) Diagram of armature winding of a two-phase generator. Black and white elements correspond to different phases. (b) A model of three-coil group similar to a one-phase group shown in (a). The direction of the magnetic field is indicated.

**Diamond-shaped Armature**

Figure 6(a) shows a textbook diagram of a two-phase generator with winding distributed in three slots per phase per pole[13]. In Fig. 6(b) a model which consists of three turns similar to each of the group in Fig. 6(a) is shown. The conductor is of the meander type used previously to make models shown in Figs. 4 and 5. In this case the twist pitch that determines the coupling loss is close to the active length of the armature. The active length of the conductor is placed into the slots edgewise, so as to minimize the component of the applied field perpendicular to the wide face.

**SUMMARY**

We have presented a novel approach to accomplishing a bending twist of the tape-like conductors similar to the $2^{nd}$ generation YBCO coated conductors. The construction of both dc field coils and ac transformer coils, as well as the superconducting stator windings may benefit from this approach, which is based on a different way of cutting the wide sheets of coated conductors into narrow tapes illustrated in Figs. 1 - 3. Although the illustrations given here are simple, they indicate the potential of new winding configurations based on coated conductor technology.

**ACKNOWLEDGEMENTS**

One of the authors, G.A.L, was supported by the National Research Council Senior Research Associateship Award at the Air Force Research Laboratory. We thank John Murphy and Jeffrey Roe for technical assistance.